# PRIVATE DELIVERY NETWORKS

*Extended Abstract*
*Presented at The Workshop on Obfuscation, May 7, 2021*

**Alex Berke, Nicolas Lee, Patrick Chwalek**
MIT Media Lab

**INTRODUCTION**
The past decade has seen tremendous shifts in how people live, work, and buy goods, with an increased reliance on e-commerce and deliveries. Purchase histories generated through e-commerce can be highly personal, revealing identifying information about individuals and households. Constructing profiles from these data allows for the targeting of individuals and communities through practices such as targeted marketing and information campaigns. Furthermore, when purchase profiles are connected with delivery addresses, these data can measure the demographics of a local community and allow for individualized targeting to reach beyond the digital realm to the physical one.

Events that accelerated shifts towards e-commerce, such as an infectious disease epidemic, have also widened equity gaps [1, 2, 3]. This work is about alternative e-commerce delivery network models that address both rising privacy and wealth inequality concerns. This includes strategies that mask and add noise to purchase histories, and allow people to "buy privacy" through charitable contributions.

**PRIVACY MODEL**
When making e-commerce transactions, customers can use a VPN, or another identity-masking tool, in order to obfuscate direct links between their identity, location, and purchase. However, simple attempts to anonymize customer transactions are not sufficient to ensure privacy. In 2015, De Montjoye et al. showed that purchase histories are so unique to individuals that even when credit card transactions are anonymized, users can be re-identified [4].

We consider knowledge of the following as privacy risks:
1. Customer profile learned through digital transaction history
2. Customer physical location

Moreover, exposed associations between (1) digital transaction history and (2) physical location present additional threats — partial privacy may be achieved if the relationship between the two is concealed. Recent years have seen increased focus on cryptocurrencies, and payment processes are rapidly evolving. For these reasons, we leave payments out of scope.

Privacy is leaked when people explicitly tell an entity, such as an online vendor, the items they are ordering, and/or delivery location. Entities can also learn this information by observing which goods are delivered to whom. Any private delivery system must assume that deliveries in and out of physical locations are observed. However, even when a package is tracked from one location to another, the full path can be kept private when routed through intermediaries. This is similar to how VPNs obfuscate routes of digital packets.

# PRIVATE DELIVERY NETWORKS: ALTERNATIVE APPROACHES

## The Delivery Private Network (DPN)
Simple DPNs work as follows:
- Customers specify a DPN as their delivery location when making their purchase.
- Vendors deliver many packages from different customers to a DPN.
- DPNs wrap received packages in additional packaging to de-identify packages.
- Upon receiving X packages of common size or after a randomized artificial delay, DPNs forward packages to customers' final locations.

This method makes it difficult for vendors or outside observers to track which orders delivered to the DPN were forwarded to which customers' locations, protecting against traffic analysis or correlation attacks.

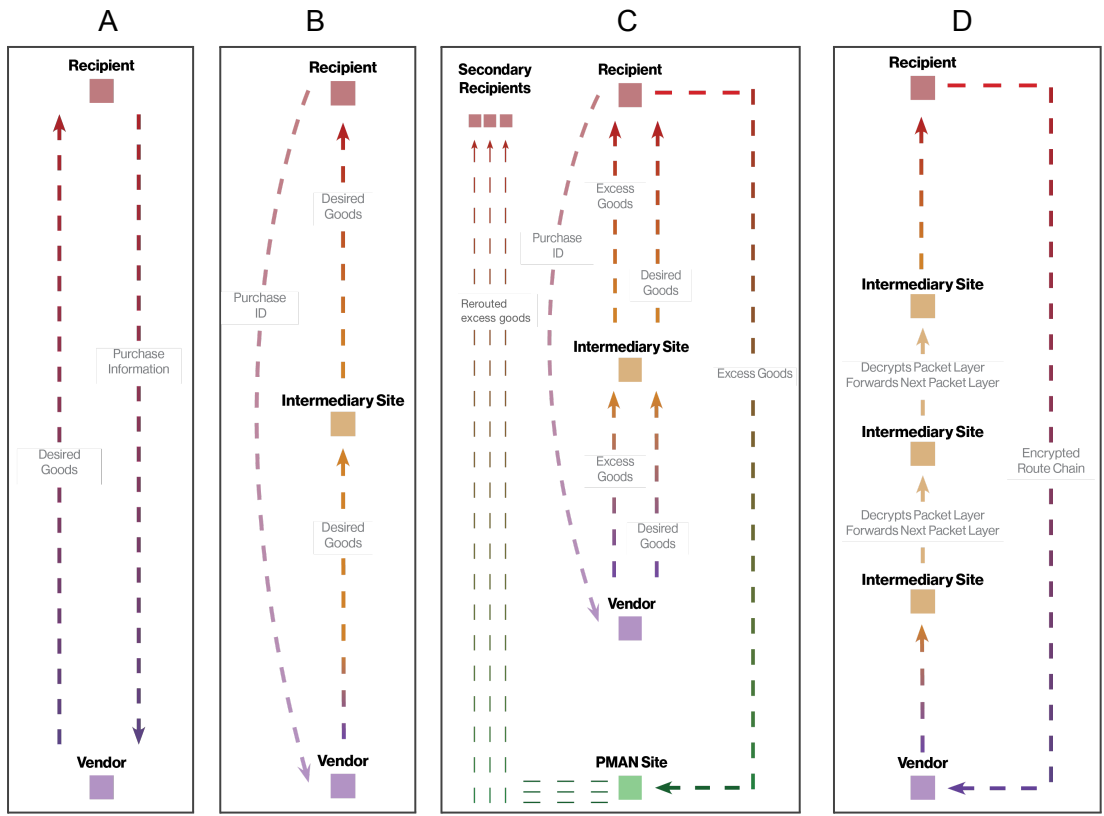

**Figure 1:** (A) Conventional e-commerce delivery network. (B) Delivery Private Network (DPN). (C) Delivery Private Network with Private Mutual Aid Network (DPN+PMAN). (D) Distributed Private Delivery Network (DPDN).

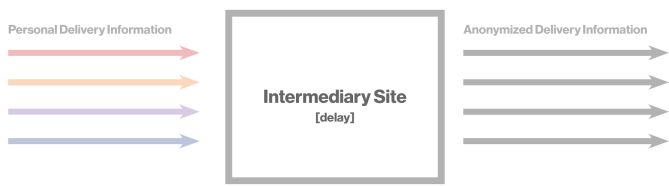

**Figure 2:** DPNs obfuscate the routes of packages by acting as intermediaries between delivery source and destination. They make it difficult for outside observers to detect which packages sent to a DPN are then redelivered where by wrapping received packages in identical packaging and forwarding them to their final destination after a privacy-enhancing delay.

A basic assumption for the simple DPN model is then that the vendor knows customer purchases but not delivery locations, while a DPN knows customer delivery locations but not what they purchased. Assuming the package de-identification process isn't compromised, DPNs provide partial privacy by removing an explicit link between customer purchases and delivery locations. However, a crucial issue is that DPNs offload privacy concerns from vendors to DPNs. DPN service providers could sniff packages, providing them with links between delivery contents and locations that they were designed to obfuscate.

**Coupling the Delivery Private Network with Noise and Wealth Redistribution: Delivery Private Networks + Private Mutual Aid Networks (DPN + PMAN)**

DPN + PMAN networks build upon the simple DPN model by adding "Private Mutual Aid Networks" (PMANs) as an additional entity in the network. DPN + PMAN networks further obfuscate user purchase data by artificially adding "noise" to purchases while incentivizing wealth redistribution: users can "buy privacy" by buying excess goods, and a PMAN ultimately redistributes those goods to secondary recipients (Figure 1.C). Users gain additional privacy from vendors as well as DPNs, who are not told which deliveries contain excess goods, or which goods within a delivery are for customers versus excess. User privacy is also protected from PMANs. Upon receiving goods, PMANs only learn what customers *did not* order for themselves - therefore learning nothing about their personal purchasing profile. PMANs also have limited knowledge of the secondary recipients' profiles since they only learn what these secondary recipients requested. Breaking the delivery network architecture into 3 separate entities - vendor, DPN, PMAN - which each have separate pieces of knowledge, further protects recipient privacy. However, this model still suffers from the privacy threat of their potential collusion and knowledge sharing.

**Distributed Private Delivery Networks (DPDN)**

If DPNs are similar to VPNs, DPDNs are more like TOR [5]. They use a series of forwarding locations (e.g. DPNs) to obscure a complete route from any one entity. Chaum laid important groundwork for these system architectures, proposing public-key cryptography as a means to obscure multi-hop routes between senders and receivers of digital packages, despite transmission occurring on unsecured links [6]. In 2016, Ike et al. proposed a system for e-commerce based on Chaum's design [7]. However, their work was limited to digital goods that could be delivered online, such as e-books and music. For physical goods, others have proposed using blockchain-based technology [8] or blind group signature schemes [9, 10, 11], instead of using encrypted layers for routing.

Chaum's original description of privacy created by each intermediary in his digital (mix) network lends itself well to the privacy created at each intermediary in a DPDN: they 'mix' packages, making it difficult for outside observers to determine which received packages were forwarded where. Only partial routes between one intermediary and the next can be observed. Even intermediaries only learn which sites come directly before or after them in a route. When there are sufficient intermediaries in a route, the complete route is obscured for all entities involved.

**DISCUSSION**

Each of the delivery network architectures presented trade efficiency for privacy by introducing intermediaries. While intermediaries can increase privacy, they incur additional delivery costs and latency. Privately delivering digital packets may be done at a relatively low cost compared to secure, private delivery of physical packages. And until personal privacy becomes a primary concern for consumers, delivery methods may be left to the mechanics of a cost-minimizing market. We use the DPN+PMAN model to look towards future markets for delivering goods and privacy — markets that can incentivize individuals who have the means and desire to buy privacy to do so by contributing to those with less means to pay for these assets.